\documentstyle[preprint,tighten,aps]{revtex}

\renewcommand{\arraystretch}{1.4}

\begin{document}


\draft
\preprint{HUB-EP-98-25, DAMTP-98-27, hep-th/9804159}
\date{April 1998}
\title{
Entropy  
and Conformal Field Theories of ${\bf AdS_3}$
Models\footnote{Research supported
by Deutsche Forschungsgemeinschaft (DFG).}}
\author{Klaus Behrndt$^1$, Ilka Brunner$^1$ and Ingo Gaida$^2$\\ ~ }

\address{$^1$Humboldt Universit\"at Berlin, Institut f\"ur Physik, \\
Invalidenstrasse 110, 10115 Berlin, Germany.\\
$^2$DAMTP, University of Cambridge\\
Silver Street, Cambridge CB3 9EW, UK}
\maketitle\

\begin{abstract}
We present a  
derivation of the Bekenstein-Hawking entropy from conformal field
theories. In particular we consider a six-dimensional string
configuration with background metric
$AdS_3 \times S^3$ near the horizon.
In addition we introduce momentum modes along the
string, corresponding to a Banados-Teitelboim-Zanelli (BTZ) black hole in the
anti-de Sitter ($AdS$) 
space-time, and a Taub-NUT soliton in the transverse Euclidean
space, projecting out a discrete subgroub of $S^3$. 
This spherical part is described
by a $\frac{SU(2)}{Z(m)}$ Wess-Zumino-Witten (WZW) model. 
The $AdS_3$ space-time, on the other hand,
is determined by two conformal field
theories living at the two boundaries:
One $\frac{SL(2,{\bf R})}{U(1)}$ WZW model on the horizon of the BTZ black
hole and one Liouville model at infinity. The
extremal BTZ black hole interpolates between these two conformal
field theories. Moreover, we argue that the sum of the three
conformal field theories yields the correct microscopic
state-counting including all $\alpha^\prime$ corrections.  
\\
\end{abstract}



%
%


\newpage

\setcounter{page}{1}
\setcounter{section}{1}

Recently there has been a lot of interest in certain $d$-dimensional
conformal field theories that can be described in terms of
supergravity and/or string theory on the product of a
($d+1$)-dimensional anti-de Sitter ($AdS$) space with a compact
manifold \cite{Mal2,Andy1,gkp,Skenderis,Ed}. In this letter we use this
property to derive from the near-horizon geometry of the
six-dimensional fundamental string solution an exact formula for the
extremal Bekenstein-Hawking entropy  from 
conformal field theories (CFTs) including all
$\alpha^\prime$ corrections.
\\
As our ``master model'' we consider the string configuration in six
dimensions. Using $U$-duality we transform all charges into $NS$-charges
and focus on four non-trivial charges. 
The general string solution has the following background metric
in the string frame \cite{mirjam}
\begin{eqnarray} \label{metric}
ds^2 = {1 \over H_1} ( du dv + H_0 du^2 ) + H_2 \left( {1 \over H_3}
(dx_4 + \vec{V} d\vec{x})^2 + H_3 \, d\vec{x} d\vec{x} \right) 
\end{eqnarray}
with $v,u = z \pm t $ and 
$\vec\nabla \times \vec{V}= \vec\nabla H_3. $ Every harmonic function 
parametrizes one brane:\\
\\
$H_1= 1 + {q_1 \over r}$: the fundamental string \newline
$H_0= 1 + {q_0 \over r}$: wave-modes along the string \newline
$H_2= 1 + {p^2 \over r}$: (compactified) NS-5-brane \newline
$H_3= 1 + {p^3 \over r}$: KK-monopole (Taub-NUT space) \newline
\\
Introducing polar coordinates, we find for the near-horizon geometry 
\begin{eqnarray} \label{6d-nearhor}
ds^2 &=& {r \over q_1} \left[ -dt^2 + dz^2 + {q_0 \over r} (dz - dt)^2 \right] 
+ p^2 p^3 \left({dr \over r}\right)^2 + 
\nonumber\\ 
& & {p^2 p^3} \left[ (d\zeta + (\pm 1-\cos\theta) \, d\phi)^2 +  
 d\Omega_2  \right] 
\end{eqnarray}
where $x^4 = p^3 \zeta$ and the ``$\pm $'' ambiguity indicates the
different choices for $V$ for the north and south hemisphere.
Thus, near the horizon $r=0$ the six-dimensional space-time becomes
a product space of two three-dimensional subspaces
\begin{eqnarray}
 M_{6} &=&  {AdS}_{3} \ \times \  S^{3}/Z_m
\end{eqnarray}
The Euclidean space $S^3/Z_m$ is described by a $SU(2)/Z(m)$-WZW model
(see \cite{Lowe} and reference therein), the discrete subgroup is
projected out due to the KK-monopole ($\zeta \simeq \zeta + {4\pi
\over m}$). This model corresponds to an {\em exact} two dimensional 
conformal field theory (CFT) with level and central charge
\begin{eqnarray}
k \equiv k_{SU} = {p^2 p^3 \over \alpha'}\ , \ \ \ \ 
c_{SU} = {3k \over k + 2} -1  \ .
\end{eqnarray}
Thus, in the classical limit $k \rightarrow \infty$ (or $\alpha' 
\rightarrow 0$) one obtains $c_{SU} =2$. Since we have
a compact group manifold $k$ has to be quantized (positive integer).
The non-compact three-dimensional space-time $AdS_3$, on the other hand,
represents the extreme
Banados-Teitelboim-Zanelli (BTZ) black hole solution \cite{BTZ}
for appropriate values of the charges 
with coordinates $x^\mu = (t,z,r)$.
Starting from (\ref{metric}), we can perform the following
coordinate transformation
\begin{eqnarray}
t \rightarrow \sqrt{ 2 q_1 \over l} \; t \qquad , \quad 
z \rightarrow \sqrt{q_1 \over 2 l} \; z \qquad , \qquad 
r \rightarrow  {2 r^2 \over l} - q_0
\end{eqnarray}
to obtain the  metric:
\begin{eqnarray}
\label{btz1}
 ds^2 &=& - e^{-2V(r)} \ dt^2 + e^{2V(r)} \ dr^2
          + \Big({r\over l}\Big)^2 \ \Big( dz -\frac{l q_0 }{2 r^2} \, dt  \Big)^2 
\end{eqnarray}
with
\begin{eqnarray}
 e^{-V(r)} = \frac{r}{l} - \frac{q_0}{2r} = \frac{r^2-r_0^2}{rl} 
\quad , \quad r_0^2 = {l q_0 \over 2} \quad , \quad l^2 = 4 p^2 p^3
\end{eqnarray}
The metric (\ref{btz1}) satisfies the boundary conditions given
in \cite{Brown} and  is therefore asymptotically anti-de Sitter with
radius $l$. The horizon of the BTZ black hole is located at $r=r_0$
and solves the three-dimensional Einstein-Hilbert action
\begin{eqnarray}
  S_{EH} &=& \frac{1}{2 \kappa_3^2} \int_{AdS_3} \ d^3 x \ \sqrt{-g} \
             (R \ - \ 2 \Lambda)
\end{eqnarray}
including a negative cosmological constant $\Lambda=-1/l^2$. 
In the limit $q_0 \rightarrow 0$ one obtains the empty space solution
($AdS$ vacuum state) with metric
\begin{eqnarray}
 ds_{\rm vac}^2 &=&  - \frac{r^2}{l^2} dt^2 +  \frac{l^2}{r^2} \ dr^2
          + \frac{r^2}{l^2} \  dz^2 \ .
\end{eqnarray}
In the context of our ``master model'' this vacuum solution is obtained
if there are no wave-modes along the six-dimensional string
configuration. Note that the vacuum solution and the ``standard''
$AdS_3$ metric
\begin{eqnarray}
 ds_{\rm AdS_3}^2 &=& - \left ( \frac{r^2}{l^2} + 1 \right ) \ dt^2 
+ \left ( \frac{r^2}{l^2} + 1 \right )^{-1} \ dr^2
          + \frac{r^2}{l^2} \  dz^2.
\end{eqnarray}
are globally inequivalent (for a detailed discussion see \cite{BTZ,Andy1}).
The BTZ black hole can be parametrised by a dreibein $e$ 
(with components $e_\mu^{ \ a}$). Here we choose the following
dreibein $e$ with inverse $e^{-1}$ (with the components $e^\mu_{ \ a}$)
\begin{eqnarray}
e = 
\left (
   \begin{array}{ccc}
   e^{-V}     & -r_0^2/rl   & 0   \\
   0          & r/l  &  0  \\
   0          & 0    & e^V \\
\end{array}
\right ), \ \ \ \
e^{-1} = 
\left (
   \begin{array}{ccc}
   e^{V}     & 0   & 0      \\
   e^V\, r_0^2 /r^2\;     & l/r &  0     \\
   0         & 0   & e^{-V} \\
\end{array}
\right ).
\end{eqnarray}
It is known that gravity in $2+1$ dimensions 
is equivalent to Chern-Simons theory \cite{Achucarro,Witten1},
where the gauge group is the $AdS$ diffeomorphism group
$SO(2,2) \sim SL(2,{\bf R})_R \times SL(2,{\bf R})_L$,
i.e. 
\begin{eqnarray}
S_{EH} &=& S_{CS} [A] \ - \  S_{CS} [\bar A]
\end{eqnarray}
with
\begin{eqnarray}
\label{CSA}
  S_{CS} [A] &=& \frac{k}{4 \pi} \int_{M_3} d^3 x 
  \ \epsilon^{\mu\nu\rho} \ {\rm Tr} \ 
  (A_\mu \partial_\nu A_\rho + \frac{2}{3} A_\mu A_\nu A_\rho )
\end{eqnarray}
Strictly speaking, this form of the Chern-Simons action
is only valid for models on a closed three-manifold $M_3$.
Here we consider $M_3 = {\bf R} \times \Sigma$, where ${\bf R}$
corresponds to the time of the covering space of $AdS_3$ and $\Sigma$
represents an ``annulus'' $r_0<r< \infty$.
In order to obtain a well-defined Chern-Simons action additional
boundary terms must be introduced.
These boundary terms have an important
consequence. Chern-Simons theory on a closed manifold
is purely topological, which is easily seen from (\ref{CSA}),
since it does not require a particular metric.
Furthermore, connections $A$ which differ only by a gauge 
transformation lead to the same contribution to the path integral.
However, introducing boundaries ``would-be'' gauge degrees of freedom
become dynamical at the boundaries.
The gauge connections of the Chern-Simons
action are related to the Einstein-Hilbert action
by \cite{Achucarro}
\begin{eqnarray}
A = (\omega^a + \frac{1}{l} e^a) \ T_a, \hspace{1cm} 
\bar A = ( \omega^a - \frac{1}{l} e^a) \ \bar{T}_a.
\end{eqnarray}
Here $\omega^a = \frac{1}{2} \epsilon^{abc} \omega_{bc}$, where
$\omega_{bc}$ denotes the spin-connection one-form. Moreover,
($T_a, \bar T_a$) are the generators of two $SL(2, \bf R)$ Lie algebras
with
\begin{eqnarray}
T_0 = \frac{1}{2} \ 
\left (
   \begin{array}{cc}
   0     & -1   \\
   1     & 0  \\
\end{array}
\right ), \ \ \ \
T_1 = \frac{1}{2} \ 
\left (
   \begin{array}{cc}
   0     & 1   \\
   1     & 0  \\
\end{array}
\right ), \ \ \ \
T_2 = \frac{1}{2} \ 
\left (
   \begin{array}{cc}
   1    & 0   \\
   0    & -1  \\
\end{array}
\right ).
\end{eqnarray}
satisfying
\begin{eqnarray}
 [ T_a , T_b ] = \epsilon_{ab}^{ \ \  c} T_c, \hspace{0,5cm} 
 [ T_a , \bar T_b ] = 0, \hspace{0,5cm}
 {\rm Tr} (T_a T_b) = \frac{1}{2} \eta_{ab}  
\end{eqnarray}
with $\eta= {\rm diag}(-1,1,1)$ and $\epsilon^{012}=1$.
The spin connections are obtained by solving
$$
de^a + \omega^a_{\quad b} \wedge e^b = 0
$$
We choose conventions where the
three-dimensional gravitational coupling is related to the
level $k$ by  
\begin{eqnarray}
 k &=& \frac{2 \pi l }{\kappa^2_3} = {p^2 p^3 \over \alpha'}.
\end{eqnarray}
After some tedious calculations one obtains
for the gauge connections
coming from the extremal BTZ black hole
\begin{eqnarray} \label{connection}
\begin{array}{ll}
  A^0 = \frac{r^2 - r_0^2}{rl^2} \ dv, \hspace{1cm}     &
  \bar A^0 = \frac{r^2 - r_0^2}{rl^2} \ du,      \\
   A^1 = \frac{r^2 - r_0^2}{rl^2} \ dv,\hspace{1cm}     &
  \bar A^1 = - \frac{r^2 + r_0^2}{rl^2} \ du,    \\
   A^2 = \frac{r^2 + r_0^2}{r^2 - r_0^2} \ \frac{dr}{r}, \hspace{1cm} &
   \bar A^2 = -  \frac{dr}{r}.    \\
\end{array}
\end{eqnarray}
or, equivalently, with $T_{\pm} = T_1 \pm T_0$
\begin{eqnarray} 
\begin{array}{l}
  A = \frac{r^2 - r_0^2}{rl^2} \ dv\ T_+ \  + \  
 \frac{r^2 + r_0^2}{r^2 - r_0^2} \ \frac{dr}{r}\ T_2  \ , \\
\bar A = \big( \frac{r^2 - r_0^2}{rl^2} \ \bar T_0 
 - \frac{r^2 + r_0^2}{rl^2} \ \bar T_1 \big) \, du - \frac{dr}{r}\ \bar T_2 \ .
\end{array}
\end{eqnarray}
It can be checked that the field strength $F=dA + A \wedge A$ as well 
as $\bar{F}$ vanish and, therefore, these connections are pure gauge
\begin{equation}
A = g^{-1} d g \qquad , \qquad \bar A = \bar g^{-1} d \bar g
\end{equation}
with $g \in SL(2, {\bf R})_R$ and $\bar g \in SL(2, {\bf R})_L$.
To understand this  from a different point of view,
let us take
the chirality conditions $ A_u = \bar A_v = 0$ as boundary
conditions. They can be implemented consistently by adding an appropriate
term to the Chern Simons action. The variation of this additional boundary term
has to cancel the boundary term coming from the variation of
the Chern-Simons action, such that the variation of the total
action does not receive corrections from the boundary.
In the Chern Simons action, the temporal components of the gauge
connection
$(A_t , \bar A_t)$ appear as Lagrange multipliers enforcing the constraint
$F_{rz} = \bar F_{rz} = 0$ (see e.g.\ \cite{Seiberg}). 
This constraint is satisfied by the extremal BTZ black hole
solution. Moreover, since in our particular model
the gauge fields depend only on the
coordinate $r$, all other field strength components vanish, too.
\\
It follows that
the Chern-Simons model or, equivalently, the extremal BTZ black hole 
can be rewritten as a WZW-model \cite{Witten3}
\begin{eqnarray}
 S_{WZW} [g] &=& 
\frac{k}{8 \pi} \int_{\partial M_3} d^2 x 
  \  {\rm Tr} \ \big[ (g^{-1} \partial_\mu g )(g^{-1} \partial^\mu g)\big] + 
\frac{k}{12 \pi} \Gamma [g]
\nonumber\\
\Gamma [g] &=& \int_{M_3} d^3 x \ \epsilon^{\mu \nu \rho}
  \  {\rm Tr} \ \big[(g^{-1} \partial_\mu g)
                (g^{-1} \partial_\nu g)
                (g^{-1} \partial_\rho g)\big] \ .
\end{eqnarray} 
where $g \in SL(2, {\bf R})$ depends only on the boundary coordinates.
$g$ is now dynamical variable. However, not all gauge
degrees of freedom are dynamical. We are still left with
a residual gauge invariance. As a consequence we have to consider
a coset model rather than a full $SL(2,{\bf R})$ WZW model.
The particular form of the coset depends on the boundary
considered (see also \cite{lee}).
In our model we have two boundaries, one at
the horizon of the BTZ black hole
($r=r_0$) and another one at infinity. Both boundaries 
are parameterized by $(u,v)$. The non-vanishing gauge fields 
$(A_v , \bar A_u)$ become
\begin{equation} \label{bound-cond}
\begin{array}{c|c}
{\rm on \ the\ horizon:}\  r= r_0 \qquad & {\rm at\ infinity:} \ r 
\rightarrow \infty \\ \hline
A_v = 0  \quad , \quad
\bar A_u = {-2 r_0 \over l^2} \, \bar T_1 \quad &  
\quad A_v = {r \over l^2} \, T_+ \ , \ 
\bar A_u = {r \over l^2} \, \bar T_- 
\end{array}
\end{equation} 
To fix all gauge degrees of freedom, we have to fix
the $T_1$-component of $\bar A_u$ at the boundary at
$r=r_0$ and the $T_+$ component
of $A_v$ and the $\bar T_-$ component of
$\bar A_u$ at the boundary at $r \to \infty$.
Alternatively, one can implement the
correct boundary conditions directly in the
WZW model. First, one adds up both chiral WZW-models to one non-chiral
$SL(2, {\bf R})$ model. Next, since the gauge fields are pure gauge,
the boundary conditions (\ref{bound-cond}) become constraints
of the $SL(2, {\bf R})$-currents of the (non-chiral) WZW-model.
These constraints can be taken into account
by gauging the corresponding group direction. 
Note that one has two {\em different}
boundary conditions at the two boundaries. 
It follows that there are also two
{\em different} CFT's at each boundary. 
\\
{\em (i) The outer boundary.} Here, the boundary conditions
(\ref{bound-cond}) yield constraints on the currents of the
WZW-model: 
\begin{equation}
 J^+_v = - {k \over 2} {\rm Tr}(T_+ g^{-1} \partial_v g)= \mbox{const},
\hspace{1cm}
 J^-_u = - {k \over 2} {\rm Tr}(T_- g^{-1} \partial_u g)= \mbox{const}. 
\end{equation}
Imposing these constraints, the $SL(2, {\bf R})$ is truncated
to a Liouville model \cite{Polyakov,Wipf,Verlinde,Driel} with 
central charge
\begin{equation}
c_{L} = {3k \over k -2} - 2 + 6k \ .
\end{equation}
This truncation corresponds to a gauged WZW-model, where a lightlike
direction has been gauged, i.e.\ to a ${SL(2, {\bf R}) \over O(1,1)}$
WZW model.
\\
{\em (ii) The inner boundary.} Here, the boundary condition does not
fix a lightlike group direction, but a spacelike. So we have to
consider a ${SL(2, {\bf R}) \over U(1)}$ WZW model, where the $U(1)$
corresponds to the $T_1$ direction.  The central charge in this case
is
\begin{equation}
c_{SL} = {3k \over k-2} -1
\end{equation}
Note that this is not the ``standard'' gauged WZW model, instead the
non-vanishing boundary conditions yield non-vanishing but fixed WZW currents.
\\
Putting it all together
the six-dimensional model
(\ref{6d-nearhor}) is determined by a superposition of three conformal
field theories.
Two of them are living on each boundary of $AdS_3$ and one
takes the $S_3/Z_m$ space into account. In order to obtain the total
central charge we must add up all contributions and obtain
\begin{eqnarray}
 c_{tot} = c_{SU} + c_{SL} + c_{L} 
         = \Big({3k \over k+2 } -1\Big) +  \Big({3k \over k-2} -1\Big) +
\Big({3k \over k-2} -2 + 6k\Big).
\end{eqnarray}
Thus, the complete entropy is given
\begin{eqnarray}
 S &=& 2 \pi \ \sqrt{\frac{1}{6} c_{tot} N} \ .
\end{eqnarray}
To compare this result with the Bekenstein-Hawking
entropy, we will embed our model in the heterotic superstring
theory. Thus, we will consider the corresponding Super-CFT.
The corresponding modifications are straightforward: In
the $SU(2)$ model we have to replace  $k$ by $k-2$  and 
in the $SL(2)$ models by $k+2$. In addition one has to add
1/2 for every fermionic mode. It follows that the total 
superconformal central charge has the general form
\begin{eqnarray}
 c_{tot} &=& 6 k \ + \ \beta \ + \ \frac{\gamma}{k},
\end{eqnarray}
Note, however, that the total central charge of
our master model, for which $\gamma= 6$,  
has a $k \leftrightarrow 1/k$ 
symmetry and, therefore, there is a lower bound for $k$: 
$c_{tot}$ becomes
minimal at $k=1$. It is interesting to note, that in the 
case of double-extreme black holes (i.e.\ for constant scalar fields)
this bound yields also a  bound
for the BPS-mass. 
In the classical limit the double-extreme BPS mass vanishes for
vanishing charges  and the classical
black hole solution ``disappears''. However,
our result indicates, 
that one obtains the lowest possible
value of the mass 
for $k=1$, corresponding to $p^2 p^3 = \alpha'$.
This is an interesting result and deserves further
investigations.
\\
The oscillator number $N$ can be obtained from the level matching
condition.
The mass of elementary heterotic string states are given by
\begin{eqnarray}
M^2_{BPS} \sim N_L -1 + \frac{1}{2} p_L^2 =
\frac{1}{2} p_R^2 = \frac{1}{4} ( \frac{q_0}{R} + q_1 R )^2
\end{eqnarray}
Here, $N_{L} \geq 0$ are number operators and $(p_L, p_R)$ is a vector in
the Narain lattice and, in terms of the integer valued
momentum [winding] quantum number $q_0$ [$-q_1$], 
$p_{L,R} = \frac{1}{\sqrt{2}}  ( \frac{q_0}{R} \mp {q_1 R  \over \alpha'} )$.
Thus, one obtains
\begin{equation}
N=N_L= 1+{q_0 q_1 \over \alpha'} \ .
\end{equation}
It follows that the entropy of the string
configuration is given in the BPS limit by
\begin{eqnarray}
 S &=& 
2 \pi \sqrt{\frac{1}{6}
\Big(1 + {q_0 q_1\over \alpha'}\Big)\Big(6k + \beta + {\gamma \over k}\Big)}
\end{eqnarray}
with $k = {p^2 p^3 \over \alpha'}$. Here $\beta$ and $\gamma$
are fixed by the fermionic contributions.
\\
Hence we found the exact microscopic form for the entropy of the 
six-dimensional string
configuration in the BPS limit including {\em all} $\alpha^\prime$ 
corrections. Compactification to four dimensions
yields the microscopic form for the corresponding black hole 
configurations. For a detailed discussion we also refer to
\cite{Skenderis}.
\\
In the limit of large $q_0$ our result
should yield the Maldacena-Strominger-Witten formula \cite{Mal1,Sabra}
\begin{eqnarray}
\label{MSW}
 S &=& 2 \pi \ \sqrt{\frac{q_0 (6 D + c_{2A} p^A) }{6}}.
\end{eqnarray}
if $D = p^1 p^2 p^3$ and $c_{2A} = c_{21} = \beta$.
Indeed, in the large $q_0$ limit we obtain $N_L = {q_0 q_1 \over \alpha'}$
and, performing the symplectic
transformation $q_1 \rightarrow p^1$ in order
to map the type II result to the
heterotic side, both results coincide for large $k$. 
Moreover, in the classical limit $k \rightarrow \infty$ (or $\alpha' 
\rightarrow 0$) one obtains
\begin{eqnarray}
 S_{\rm classical} &=& 2 \pi \ \sqrt{q_0 D}, \ \ \ \
c_{\rm classical} = 6k.
\end{eqnarray}
It follows that the classical entropy is given by large $N_L$ and
the central charge of the Liouville model, only. Note that this
result holds for supersymmetric and non-supersymmetric central
charge of the Liouville model.
\\
Moreover, it is straightforward to generalize the results to
an $O(m,n)$ invariant form (see \cite{Ingo1} for example): 
\begin{eqnarray}
 q_0 q_1 &\rightarrow& \vec Q L \vec Q,
 \hspace{1cm} 
 p^2 p^3 \ \rightarrow \ \vec P L \vec P,
 \hspace{1cm} 
  \vec P,\vec Q \in O(m,n).
\end{eqnarray}
Here $L$ denotes the $O(m,n)$ metric. Thus, one obtains
\begin{eqnarray}
 N_L &=& \vec Q L \vec Q + 1, \ \ \mbox{and} \ \
 k =  {\vec P L \vec P \over \alpha'}.
\end{eqnarray}
In summary, we have shown that the six-dimensional string
configuration yields a product space near the horizon $r=0$.
This product space yields four WZW models at the boundaries,
which can be reformulated as three different conformal field
theories. This result can be used to find the relevant microscopic
quantum states giving a statistical interpretation to the
macroscopic Bekenstein-Hawking entropy. Here, we computed the exact
entropy in the BPS limit including
all $\alpha^\prime$ corrections from CFTs. 
One important result
of this letter is that we found one
$\frac{SL(2,{\bf R})}{U(1)}$ gauged WZW model
at the inner boundary $B_0$, i.e.
at the horizon of the BTZ black hole, and 
two chiral WZW models at the outer boundary $B_\infty$
giving rise to one
Liouville model \cite{Driel,lee}. 
The extreme BTZ black hole interpolates between these two boundaries
with their CFTs. 
We argued that the
sum of all CFTs,
living at the boundaries of the
near-horizon space-time, 
yields the correct microscopic state-counting procedure.
Thus, we follow the general idea that the Bekenstein-Hawking
entropy should be accounted
for by microstates near the horizon \cite{Hor,Andy1}.
Note that we only considered the leading order contributions to
the Bekenstein-Hawking entropy. In general there are further
subleading (logarithmic) contributions \cite{IB}.
\\
One should keep in mind, that our total central charge is not fixed by
the total central charge of the conformal anomaly of
heterotic superstring theory. As pointed out before the crucial point is 
that we discussed {\em only boundary} CFT's. Since the Liouville field
parametrises radial fluctuations of $AdS_3$ it resembles very much
the old disussion that non-critical string theory in $d$ dimensions
becomes critical in $d+1$ dimensions.
In this context the Liouville field
compensates for the $d$-dimensional matter central charge. For
a discussion on these issues we refer also to \cite{gkp}.  
\\
This letter motivates the following picture on
the location of the microscopic 
states that are encoded in the four-dimensional
Bekenstein-Hawking entropy. Classically ($k \rightarrow \infty$)
the only contribution to the entropy comes from the
Liouville model living at the outer $AdS$ boundary. It is
suggestive to understand this boundary as the mouth region
of the black hole throat geometry. But this is not the whole
story, since additional
subleading contributions come from states located at
the bottom of the throat, which are independent of the Liouville
mode.
\\
It would be very interesting to obtain additional macroscopic 
examples including all $\alpha^\prime$ corrections.
In fact, currently
we prepare another article on the
general setup presented in this letter including further details and
extensions.

\bigskip  \bigskip

\noindent
{\bf Acknowledgments}  \medskip \newline
We would like to thank A. Tseytlin, T. Mohaupt, S. F\"orste, T. Turgut
and M. Walton for discussions.

\bigskip

\noindent
{\bf Note added:} \newline
While writing this letter we received \cite{Mal3,Mar} which have some
overlap with aspects of our work.

%
%
%

\renewcommand{\arraystretch}{1}

\newcommand{\NP}[3]{{ Nucl. Phys.} {\bf #1} {(19#2)} {#3}}
\newcommand{\PL}[3]{{ Phys. Lett.} {\bf #1} {(19#2)} {#3}}
\newcommand{\PRL}[3]{{ Phys. Rev. Lett.} {\bf #1} {(19#2)} {#3}}
\newcommand{\PR}[3]{{ Phys. Rev.} {\bf #1} {(19#2)} {#3}}
\newcommand{\IJ}[3]{{ Int. Jour. Mod. Phys.} {\bf #1} {(19#2)}
  {#3}}
\newcommand{\CMP}[3]{{ Comm. Math. Phys.} {\bf #1} {(19#2)} {#3}}
\newcommand{\PRp} [3]{{ Phys. Rep.} {\bf #1} {(19#2)} {#3}}


\begin{thebibliography}{9}

\bibitem{Mal2}
P. Claus, R. Kallosh, A. Van Proeyen, {\tt hep-th 9711161};
\\
J. Maldacena,  {\tt hep-th 9711200};
\\
S.-J. Rey, S. Theisen and J.-T. Yee, {\tt hep-th 9803135}; 
\\
H. Liu and A.A. Tseytlin, {\tt hep-th 9804083}, 


\bibitem{Andy1}
A. Strominger,  {\tt hep-th 9712251}\\
N.\ Kaloper, {\tt hep-th/9804062}.

\bibitem{gkp}
S.S. Gubser, I.R. Klebanov and A.M. Polyakov,
{\tt hep-th/9802109}

\bibitem{Skenderis}
K. Sfetsos and K. Skenderis, {\tt hep-th 9711138};
\\
H.J. Boonstra, B. Peeters and K. Skenderis, {\tt hep-th 9803231}.

\bibitem{Ed}
E. Witten, {\tt hep-th 9802150}, {\tt hep-th 9803131};

\bibitem{mirjam}
M. Cvetic and D. Youm, \PR{D53}{96} 584, {\tt hep-th/9507090};
\\
M. Cvetic and A.A. Tseytlin, \PR{D53}{96} 5619, {\tt hep-th/9512031}.

\bibitem{BTZ}
M. Banados, C. Teitelboim and J. Zanelli, \PRL{69}{92}{1849};
\\
M. Banados, M. Henneaux, C. Teitelboim, J. Zanelli,
\PR{D48}{93}{1506}

\bibitem{Brown}
J.D. Brown and M. Henneaux, \CMP{104}{86}{207}.


\bibitem{Achucarro}
A. Achucarro and P.K. Townsend, \PL{B180}{86}{89}.

\bibitem{Witten1}
E. Witten,\NP{B311}{88/89}{46}. 


\bibitem{Seiberg}
S. Elitzur, G. Moore, A. Schwimmer and N. Seiberg,\NP{B326}{89}{108}.

\bibitem{Witten3}
E. Witten, \CMP{92}{84}{455}. 


\bibitem{lee} 
T. Lee, {\tt hep-th/9706174}

\bibitem{Driel}
O. Coussaert, M. Henneaux, P. van Driel, Class.Quant.Grav. {\bf 12}
(1995) 2961. 


\bibitem{Lowe}
D.A. Lowe and A. Strominger, \PRL{73}{94}{1468}.  


\bibitem{Polyakov}
A.M. Polyakov, Mod. Phys. Lett. {\bf A2} (1987) 893;
\\
V.G. Knizhnik, A.B. Zamolodchikov, A.M. Polyakov, 
Mod. Phys. Lett. {\bf A3} (1988) 819;


\bibitem{Verlinde}
R. Dijkgraaf, H. Verlinde, E. Verlinde, \NP{B371}{92}{269}. 


\bibitem{Wipf}
A. Alekseev and S. Shatashvili, \NP{B323}{89}{719}.
\\
M. Bershadski and H. Ooguri, \CMP{126}{89}{49}.
\\
P. Forgacs, A. Wipf, J. Balog, L. Feher, L. O'Raifeartaigh,
\PL{B227}{89}{214};


\bibitem{Mal1}
J. Maldacena, A. Strominger, E. Witten, JHEP {\bf 12} (1997) 002.

\bibitem{Sabra}
K. Behrndt, G.L. Cardoso, B. de Wit, D. L\"ust, T. Mohaupt,
W.A. Sabra, {\tt hep-th 9801081};

\bibitem{Ingo1}
M. Cvetic, I. Gaida, \NP{B505}{97}{291}, and reference therein.

\bibitem{Hor}
W.H. Zurek and K.S. Thorne, \PRL{54}{85}{2171};
\\
G. 't Hooft, \NP{B335}{90}{138},
\\
L. Susskind, L. Thorlacius, R. Uglum, \PR{D48}{93}{3743};
\\
S. Carlip, \PR{D51}{95}{632};

\bibitem{IB}
K. Behrndt and I. Gaida, \PL{B401}{97}{263};
\\
K. Behrndt, G.L. Cardoso and I. Gaida, \NP{B506}{97}{267},



\bibitem{Mal3}
J. Maldacena and A. Strominger, {\tt hep-th 9804085};

\bibitem{Mar}
E.J. Martinec, {\tt hep-th 9804111};


\end{thebibliography}
\end{document}